\documentclass[useAMS,usenatbib,usegraphicx]{mn2e}
\title[Search for Class I methanol masers]
  {Search for Class I methanol masers in low-mass star formation regions}
\author[S. V. Kalenskii et al.]{
  S. V. Kalenskii$^{1}$\thanks{E-mail:kalensky@asc.rssi.ru (SVK)}; 
  \fbox{L. E. B. Johansson$^{2},$} P.~Bergman$^{2}$,
  S. Kurtz$^{3}$,  P. Hofner$^{4,5,6}$, 
  \newauthor C. M. Walmsley$^{7}$, and \fbox{V. I. Slysh$^{1}$}\\
  $^{1} $Astro Space Center, Lebedev Physical Institute, 
     Profsoyuznaya 84/32, Moscow, 117997, Russia\\
  $^{2}$ Onsala Space Observatory, SE-439 92 Onsala, Sweden\\
  $^{3}$ Centro de Radioastronomia y Astrofizika,  Universidad Nacional 
  Autonoma de Mexico (Morelia, Michoacan, Mexico)\\
  $^{4}$ Physics Department, New Mexico Tech,
  801 Leroy Place, Socorro, NM 87801\\
  $^{5}$ National Radio Astronomy Observatory,
  PO Box 0, Socorro, NM 87801\\
  $^{6}$ Max-Planck-Institut f\"ur Radioastronomie,
  Auf dem H\"ugel 69, 53121, Bonn, Germany\\
  $^{7}$ Osservatorio Astrofisico di Arcetri, Largo E. Fermi 5,1-50125 Firenze,
  Italy
}
\begin{document}

\date{Accepted . Received ; in original form }

\pagerange{\pageref{firstpage}--\pageref{lastpage}} \pubyear{2009}

\maketitle

\label{firstpage}

\begin{abstract}
A survey of young bipolar outflows in regions of low-to-intermediate-mass 
star formation has been carried out in two class I methanol maser transitions:
$7_0-6_1A^+$ at 44~GHz and $4_{-1}-3_0E$ at 36~GHz.
We detected narrow features towards NGC~1333I2A, NGC~1333I4A, HH25MMS, and 
L1157 at 44~GHz, and towards NGC~2023 at 36~GHz. Flux densities of the lines
detected at 44~GHz are no higher than 11 Jy and the relevant source luminosities
are about $10^{22}$~erg~s$^{-1}$, which is much lower 
than those of strong masers in high-mass star formation regions. 
No emission was found towards 39 outflows. All masers detected at 44~GHz 
are located in clouds with methanol column densities
of the order of or larger than a few $\times 10^{14}$~cm$^{-2}$.  
The upper limits for the non-detections are typically of the order of 3--5~Jy. 
Observations in 2004, 2006, and 2008 did not reveal any significant variability 
of the 44 GHz masers in NGC~1333I4A, HH25MMS, and L1157.
\end{abstract}


\section{Introduction}

Bright and narrow maser lines of methanol (CH$_3$OH) have been found 
towards many star-forming regions~\citep{hmb,menten91a,kurtzetal}. According 
to the classification of~\citet{menten91b}, methanol masers can
be divided into two classes, I and II, with each class characterized by
a certain set of transitions. The Class I maser transitions 
are the $7_0-6_1A^+$ transition at 44~GHz, $4_{-1}-3_0E$ transition at 36~GHz, 
$5_{-1}-4_0E$ transition at 84 GHz, $8_0-7_1A^+$ transition at 95~GHz etc., 
while the Class II transitions are the $5_1-6_0A^+$ transition at 6.7~GHz, 
$2_0-3_{-1}E$ transition at 12~GHz, the $J_0-J_{-1}E$ series of
transitions at 157~GHz, etc. A list of the most powerful Class 
I and II transitions is presented 
in e.g.,~\citet{cragg}. Both Class I and Class II masers are often overlaid
upon broad thermal lines. Many methanol transitions, e.g.,
the series of $2_K-1_K$ transitions near 96~GHz, $3_K-2_K$ transitions
near 145~GHz, $5_K-4_K$ transitions near 241~GHz etc. never exhibit maser
features. These transitions are often called ``purely thermal". 

The nature of methanol  masers is still unknown. \citet{plammen} 
suggested that Class I masers arise in postshock gas in the lobes 
of bipolar outflows, where the abundance of methanol is enhanced due 
to grain mantle evaporation. This hypothesis has further support
in the fact that in a number of star-forming regions Class I masers 
appear to be associated with outflows~\citep{kurtzetal,chen}.  
However, this hypothesis is not generally accepted because 
there are no high-velocity Class I masers and the apparent association
between the masers and the outflows may be caused by the fact
that both of them arise in the same regions of star
formation rather than by a physical association between 
these objects. 

The difficulties in the exploration  of methanol masers partly 
appear because until recently the masers have been observed in regions of massive 
star formation, which are relatively distant (2--3 kpc from the Sun or farther)
and highly obscured at the optical and even NIR wavelengths. In addition, 
high mass stars usually form in clusters. These properties make it difficult 
to resolve maser spots and to associate masers with other objects in these 
regions. In contrast, regions of low-mass star formation are much more 
widespread and many of them are only
200--300 pc from the Sun; they are less heavily obscured than
regions of high-mass star formation, and there are many
isolated low-mass protostars. Therefore, the study of masers in
these regions might be more straightforward
compared to that of high-mass regions, and hence, the detection of 
Class I masers in low-mass regions might have a strong impact 
on maser exploration. Bearing this in mind, we performed in 2004 
a~``snapshot" search for Class I methanol masers towards bipolar outflows 
driven by low-mass YSOs~\citep{paperI} (Paper I) at 44, 84, and 95~GHz. 
The source list consisted of five so-called chemically rich outflows, 
where the abundances of methanol and some other molecules are increased
as a result of grain mantle evaporation. The search proved 
successful: three maser candidates, NGC~1333I4A, HH25, 
and L1157 were found at 44~GHz.
VLA observations of L1157 at 44~GHz confirmed that this source is really
a maser~\citep{kalen10}. Therefore a further work in the field looks promising.

In order to obtain a general idea about the main properties of the Class I
masers in the regions of low-mass star formation, we performed a more 
extended search for these objects. The new survey was carried out at the frequency
of the $7_0-6_1A^+$ transition, but most sources were additionally observed 
in another Class I maser line, the $4_{-1}-3_0E$ line at 36~GHz. 
The physical relation between low-mass protostellar outflows and
Class I methanol masers is poorly understood. Hence,
it is tempting to make a comprehensive survey of such outflows.
Unfortunately, the enormous amount of observing time makes such a survey 
unrealistic. The naive expectation, supported by the successful search
of ~\citet{paperI} is to find methanol masers towards bright thermal 
sources of methanol; therefore the basis of our source list consists of 
outflows where~\citet{kalen07} detected thermal emission in the $5_{-1}-4_0E$,
$8_0-7_1A^+$, and/or $2_K-1_K$ methanol lines. We included also three outflows, 
IRAS03282, Serpens S68FIRS1 and Serpens SMM4, where~\citet{bach95} 
and~\citet{garay02} found a significant enhancement of methanol abundance
relative to that in quescent gas. Because methanol enhancement has been detected
in young, well-collimated outflows from Class 0 and I sources, we 
included several such objects in our list regardless of whether methanol enhancement
had been found there.  
A subsample of our list consisted of YSOs with known outflows and/or 
H$_2$O masers located in Bok globules~\citep{yc92,cbh2o}. Like other 
objects from our list, these YSOs are typically isolated objects of low or
intermediate mass, located in nearby ($<$500~pc) small and relatively simple 
molecular clouds. In total, our source list consisted of 37 regions which harbor 
46 known outflows driven by Class 0 and I low-mass protostars, taken from the literature.
The observed sources, positions, and the relevant literature are given
in Tables~\ref{gauss} and~\ref{negative}.
 
In addition to the survey we performed second- and
third-epoch 44~GHz observations of the maser candidates detected in 2004. 

\begin{table}
\centering
\begin{minipage}{80mm}
\caption{The parameters of the observed lines and those of the OSO 20-m 
at the line frequencies. The parameters of the lines, observed in 
the preliminary survey~\citep{paperI} are included for completeness.
}
\label{freqtabl}
\begin{tabular}{|l|c|c|c|c|}
\hline
Transition       &Frequency  &$S\mu^2\;^a$& HPBW  & G \\
                 &  (GHz)    & (Debye) &$('')$    & (Jy/K) \\
\hline
$7_0-6_1A^+$     & 44.069476  & 6.1380& 82   &20.5\\
$5_{-1}-4_0E$    & 84.521206  & 3.0830& 44   &22\\
$8_0-7_1A^+$     & 95.169516  & 7.2211& 39   &25\\
$2_{-1}-1_{-1}E$ & 96.739393  & 1.2133& 39   &25\\
$2_0-1_0A^+$     & 96.741377  & 1.6171& 39   &25\\
$2_0-1_0E$       & 96.744549  & 1.6167& 39   &25\\  
$2_1-1_1E$       & 96.755507  & 1.2443& 39   &25\\
$4_{-1}-3_0E$    & 36.169290  & 2.5184& 105  &18\\
\hline
\end{tabular}

\medskip
$^a$--The product of the permanent dipole moment and the line
strength from~\cite{muller}\\
\end{minipage}
\end{table}

\begin{figure*}
\includegraphics[width=150mm]{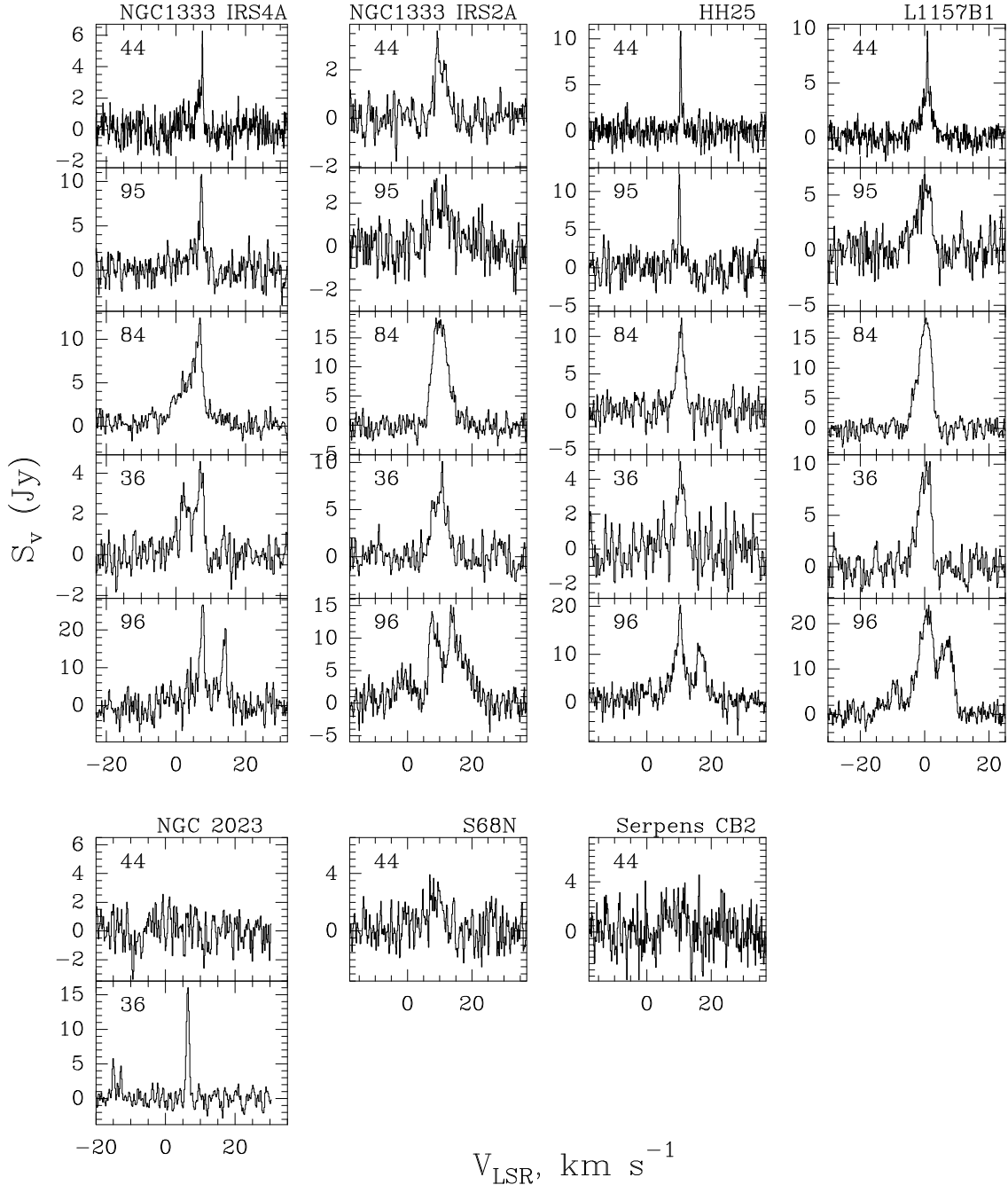}
\caption{Upper panels: spectra of the regions of low-mass star formation 
in which maser candidates in 
the $7_0-6_1A^+$ line were detected. Shown from top to bottom
are the $7_0-6_1A^+$,  $8_0-7_1A^+$, $5_{-1}-4_0E$, $4_{-1}-3_0E$,
and $2_K-1_K$ lines at 44, 95, 84, 36, and 96~GHz, respectively. 
The horizontal axis plots the radial velocity in km~s$^{-1}$ and 
the vertical axis the spectral flux density in Jansky. The 84~GHz, 95~GHz, 
and 96~GHz spectra are taken from Paper I for all sources except L1157,
for which the new spectra at all these frequencies except 96 GHz have been 
taken towards the stronger maser position.
Lower panels: spectra of other sources detected at 44 and/or 36~GHz.}
\label{mas}
\end{figure*}

\begin{table*}
\centering
\begin{minipage}{150mm}
\caption{The gaussian parameters of detected lines. Here 44 denotes 
the $7_0-6_1A^+$ at 44~GHz, 36, the $4_{-1}-3_0E$ line at 36~GHz.
For NGC~2023, the numbers in parentheses show the R.A. and DEC. offsets in
arcsec from the position given in the second and third columns.}
\label{gauss}
\begin{tabular}{|l|c|r|c|c|c|c|r|c|c|l|}
\hline
Source      &R.A.      & DEC  &Line&$\int S_{\nu}dV$    &$V_{\rm LSR}$& FWHM  &$S_{\nu}$&Obs.&Notes$^a$&Refs\\
            &(J2000)&(J2000)&$ $&(K$\cdot$km~s$^{-1}$)&(km~s$^{-1}$)&(km~s$^{-1}$)&(Jy)&(year)&&\\
\hline
NGC~1333I2A &03 29 01.0& 31 14 20 &44& 4.92(1.64)&  9.24(0.12)&1.66(0.42)& 2.81 & 2007 & r  &1,2\\ 
            &          &          &  & 5.74(1.64)& 11.57(0.31)&2.71(0.79)& 1.97 &      &    & \\ 
            &          &          &36& 13.7(1.44)&--7.10(0.37)&7.88(1.04)& 1.64 & 2007 &    & \\ 
            &          &          &  & 13.7(2.70)&  8.07(0.15)&2.00(0.26)& 6.43 &      &    & \\ 
            &          &          &  & 10.4(4.14)& 12.48(1.25)&6.72(1.36)& 1.46 &      &    & \\ 
            &          &          &  & 33.7(4.86)& 10.54(0.08)&2.49(0.25)&12.71 &      &    & \\ 
NGC~1333I4A &03 29 10.3& 31 03 13 &36& 10.6(1.98)& 2.73(0.38)& 4.61(1.22)& 2.14 & 2007 & c  &1,3\\ 
            &          &          &  & 7.56(1.44)& 7.11(0.16)& 1.97(0.34)& 3.60 &      &    & \\ 
            &          &          &44& 3.28(0.41)& 6.49(0.26)& 5.10(0.59)& 1.95 & 2007 &    & \\ 
            &          &          &  & 1.85(0.41)& 7.51(0.02)& 0.33(0.05)& 5.13 &      &    & \\ 
NGC~2023(0,0)&05 41 28.5&--02 19 19&44&          &           &          &$<3.60$& 2007 & b  & 4\\  
(0,0)       &          &          &36&15.3(0.7)  & 6.46(0.02)& 0.92(0.05)&15.73 & 2008 & b  & \\ 
(20,20)     &          &          &  & 18.5(0.82)& 6.39(0.02)& 0.99(0.05)&17.39 &      & b  & \\  
($-60,-60$) &          &          &  &           &           &           &$<3.6$&      & q  & \\ 
($60,-60$)  &          &          &  &           &           &           &$<3.0$&      & c  & \\ 
(60,60)     &          &          &  & 0.63(0.04)& 6.55(0.04)& 1.36(0.11)& 0.44 &      & q  & \\ 
($-60,60$)  &          &          &  &           &           &           &$<3.6$&      & q  & \\ 
            &          &          &  &           &           &           &      &      &    & \\ 
HH25MMS     &05 46 06.5&--00 13 54&44& 5.33(0.82)&10.51(0.04)& 0.48(0.09)& 10.41& 2007 & r;mo&1,5\\ 
            &          &          &36& 11.9(1.08)&10.56(0.12)& 2.93(0.28)& 3.82 & 2007 &    & \\ 
S68N        &18 29 47.5& 01 16 51 &44& 12.3(1.64)&  8.86(0.32)&5.18(0.83)& 2.23 & 2007 & c;mo& 6 \\ 
Serpens CB2 &18 29 58.4& 01 13 35 &44& 8.61(1.63)&  8.13(0.64)&6.10(0.87)& 1.32 & 2007 & b;mo& 7 \\ 
L1157       &20 39 08.1& 68 01 14 &44& 12.0(0.60)& 0.69(0.08) &3.82(0.24)& 3.4  & 2004 &    &1,8,\\ 
            &          &          &  & 2.40(0.20)& 0.75(0.01) &0.37(0.03)& 6.2  &      &    &9,10 \\ 
            &20 39 10.0& 68 01 42 &36& 36.6(3.60)&--0.70(0.20)&3.97(0.31)& 8.6  & 2008 & m  & \\ 
            &          &          &  & 10.8(3.20)&  1.40(0.12)&1.66(0.27)& 6.2  &      &    & \\ 
            &          &          &44& 15.1(0.57)&  0.61(0.08)&3.90(0.20)& 3.6  &      &    & \\ 
            &          &          &  &  2.0(0.32)&  0.91(0.03)&0.53(0.08)& 3.5  &      &    & \\ 
\hline
\end{tabular}
\end{minipage}
\flushleft
$^a$--r, red wing; b, blue wing; c, central position; q, quescent gas; mo, multiple outflows;
m, maser position determined with the VLA.

1--Kalenskii et al. (2007); 2--Bachiller  et al. (1998); 3--Blake et al. (1995); 4--Sandell et al. (1999);
5--Gibb \& Davis (1998); 6--Garay et al. (2002); 7--Davis et al. (1999); 8--Bachiller et al. (1995); 
9--Bachiller et al. (2001); 10--Benedettini et al. (2007); 

\end{table*}

\section{Observations}
Both the new survey and the multi-epoch observations of previously detected 
sources were carried out with the same telescope as the 2004 observations, 
namely, the 20-m radio telescope of the Onsala Space 
Observatory (OSO). The second-epoch observations were made in 
December 2006, and the new survey at 44 and 36~GHz was carried out
in December 2007. Several sources, including the three maser candidates
detected in 2004, were reobserved at Onsala in December 2008 with the same 
receiver and spectrometer setup as in 2007.
The line rest frequencies and strengths and the main telescope
parameters are presented in Table~\ref{freqtabl}.
The frequencies were taken from the Lovas
database\footnote{http://physics.nist.gov/cgi-bin/micro/table5/start.pl}.
The dual beam switching mode with a frequency of 2~Hz and a beam throw
of $11'$ was applied. Pointing errors were checked using
observations of SiO masers and were found to be within $5''$. The data
were calibrated using the chopper-wheel method. An autocorrelator 
configured to either a 12.5~kHz (0.085~km~s$^{-1}$ at 44~GHz)
or 25~kHz resolution was used as the spectrometer. An overall check of the system
was achieved by regularly observing known sources at 36 and 44~GHz.
Typically, we observed several positions per source to cover the whole area 
occupied by the outflow lobes. 

The data were reduced using the Grenoble CLASS package. 

\section{Results}
Based on the single-dish observations, we confirmed the existence
of the three maser candidates at 44~GHz, reported in Paper I, detected
broad lines at 36~GHz towards them, and found six new sources 
at 36~or 44~GHz or both. The spectra of the detected sources are shown in Fig.~\ref{mas}
and the gaussian parameters of the detected lines are presented in 
Table~\ref{gauss}; Fig~\ref{mas} shows the spectra of maser 
candidates presented in Paper I in addition to the emission detected in 
2007 or 2008. 

Only one source, newly detected at 44~GHz in 2007, NGC~1333I2A
show narrow spectral features, which may be masers. 
In the case of S68N and Serpens~CB2 the lines at 44~GHz are broad,
5--6~km~s$^{-1}$.
A fairly narrow line was detected at 36~GHz towards the blue lobe of an 
extreme high-velocity outflow in the vicinity of the bright reflection 
nebula NGC~2023. Offset measurements show
that the source is compact with respect to the 105-arcsec beam. The line has 
no counterpart at 44~GHz. Its LSR velocity, $\sim 6.4$~km~s$^{-1}$, is different
from the systemic velocity of about 10~km~s$^{-1}$~\citep{n2023}.

With the exception of NGC~2023, only broad lines were detected at 36~GHz.


Negative results are presented in Table~\ref{negative}.
No emission was detected in 39 outflows. The upper limits for 
the non-detections are typically of about 3--5~Jy. Thus, 
masers in these regions are fairly rare and/or weak objects.
It is of interest to perform a similar survey with higher sensitivity
of 1~Jy or better. 

\begin{figure}
\includegraphics[width=7cm]{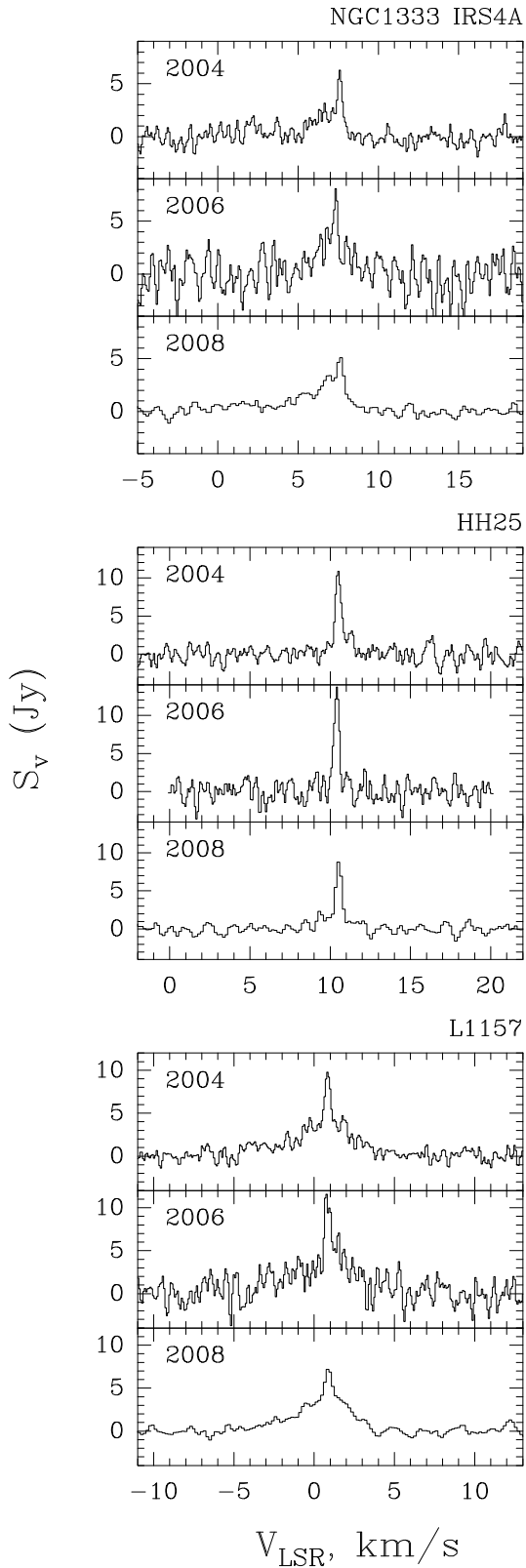}
\caption{Spectra of three maser candidates, acquired in December
2004, 2006 and 2008 in the $7_0-6_1A^+$ at 44~GHz.}
\label{0408flux}
\end{figure}

Figure~\ref{0408flux} presents the 44~GHz spectra of the three maser
candidates detected in December 2004, 2006, and 2008. The spectra do not show
notable variation between the epochs. Slight changes in line shapes
and a decrease of flux densities of all three sources in 2008, about
30\%, can be attributed to poor signal-to-noise ratios, calibration
uncertainties, and different spectral resolution (0.17~km~s$^{-1}$ in
2008 vs 0.085~km~s$^{-1}$ in 2004 and 2006). However, further
monitoring of these sources would be desirable.

\section{Discussion}
The three regions of low-mass star formation, newly detected in 2007 at 44~GHz, 
NGC~1333I2A, S68N, and Serpens CB2, exhibit weak lines ($\le 4$~Jy), making 
it difficult to determine whether they are masers. The lines are broad, 
5--6~km~s$^{-1}$, which is typical for thermal emission rather than for masers.
However, the line detected in NGC~1333I2A is poorly approximated 
by a single gaussian and a satisfactory fit is obtained with a narrow line 
overlaid on a broader component (see Table~\ref{gauss}). The lines detected 
in S68N and Serpens CB2 can be approximated by single gaussians. 
Therefore, for the present we consider the line 
detected in NGC~1333I2A to be a weak maser overlaid upon 
thermal emission, and those detected in S68N and Serpens CB2 to be
thermal lines. Poor signal-to-noise ratios in all these cases prevented
us from accurately measuring line shapes; more sensitive observations may alter 
our interpretation.

The nature of the 36~GHz line towards the blue lobe of the bipolar 
outflow in NGC~2023 is unclear. On the one hand, the line is fairly narrow, 
and offset measurements (Table~\ref{gauss}) show that the source is compact 
with respect to the 105-arcsec Onsala beam. These properties indicate that 
the source is probably a maser. This assumption has further support in the fact 
that the line LSR velocity, $\approx 6.5$~km~s$^{-1}$, is slightly less than 
the systemic velocity of about 10~km~s$^{-1}$. On the other hand, 
the line has no counterpart at 44~GHz, which is more typical for thermal
emission. Note, however, that there are known masers at 36~GHz without 44-GHz
counterparts; in particular, no 44 GHz emission was found at the velocity
of a fairly strong 36-GHz maser detected $3'$ north of DR21(OH) 
by~\citet{pratap}. Therefore we preliminary conclude that the narrow line
in NGC~2023 is a maser.

An examination of Tables~\ref{gauss} and~\ref{negative} shows
that outflows with masers at 44 GHz exhibit the strongest thermal
emission at 36 GHz among all sources from our list. As the intensities
of optically thin thermal lines are roughly proportional to column
densities, this result indicates that methanol masers arise in regions
of low-mass star formation with the highest column densities of
methanol. This conclusion can be confirmed on the basis of the results
by~\citet{kalen07} on thermal emission of methanol toward outflows
driven by low-mass YSOs. \citet{kalen07} observed thermal emission in
the $5_{-1}-4_0E$, $8_0-7_1A^+$, and a series of $2_K-1_K$ methanol
lines at 3~mm. Their angular resolution was about $40''$,
corresponding to a linear resolution of about 0.06 pc at a typical
distance of 300 pc.  Moreover, their source list was essentially a
subsample of our list. Column densities of methanol, derived for
NGC~1333I2A and I4A, HH25, and L1157 using rotational diagrams are
about $10^{15}$~cm$^{-2}$ or more.  According to the
results of LVG modeling, these column densities may be overestimated
by a factor of 3--8~\citep{kalen07}; therefore we conclude that
masers at 44 GHz can arise in regions with methanol column densities
no less than several times $10^{14}$~cm$^{-2}$. Because the
number of detected masers is small, this conclusion is not
statistically robust.


We note that the isotropic luminosities of the masers at 44 GHz in the regions 
of low-mass star formation are about $10^{22}$~erg~s$^{-1}$, i.e., several orders 
of magnitude lower than the maser luminosities in  regions of massive star 
formation. ~\citet{kalen02} observed regions of massive star formation with
a linear resolution about 0.15 pc and determined methanol column densities 
toward strong Class I masers using an approach 
similar to that of~\citet{kalen07}. They obtained values between 
$2-74 \times 10^{16}$~cm$^{-2}$. Thus, a general trend seems to be as follows:
molecular clouds with methanol column densities less than $10^{14}$~cm$^{-2}$ 
cannot produce masers at 44 GHz; clouds with methanol column densities 
$10^{14}-10^{15}$~cm$^{-2}$ can (but do not necessarily) produce weak 
masers with luminosities about $10^{22}$~erg~s$^{-1}$ ;
clouds with methanol column densities higher than $10^{16}$~cm$^{-2}$
can produce strong masers with luminosities $10^{24}-10^{25}$~erg~s$^{-1}$ .
Note that here we imply column densities averaged over fairly large regions
(about 0.06 pc in the case of low-mass regions and about 0.15 pc in the case
of high-mass regions); individual clumps inside these regions may have much
higher column densities of methanol. 

Several mechanisms of methanol maser excitation have been proposed. Most of
them probably can 
be ''scaled'' so-as to explain the existence of stronger masers only in sources 
with higher methanol column densities. Among these mechanisms are shocks driven 
by outflows, turbulence, which can result in a random increase of coherence length 
along a certain line~\citep{sam98}, and accretion shocks~\citep{kurtzetal}.
The two former mechanisms are discussed in more detail by \citet{kalen10} with respect to
the masers in L1157. Currently we cannot rule out any of these mechanisms. 
It may happen that in different sources different mechanisms are responsible for 
the maser emission or even that different mechanisms may coexist 
within the same source. 

With the exception of NGC~2023, only broad lines were detected at 36~GHz.
Earlier, \citet{kalen01} detected broad $5_{-1}-4_0E$ methanol lines at 84~GHz 
toward a number of sources and showed that these lines are typically inverted;
the fact that they are broad is a result of their low optical depths.
These lines arise in extended clouds and in all respects except excitation 
temperature are similar to thermal lines. Such lines are called quasi-thermal lines. 
The excitation of the $4_{-1}-3_0E$ lines at 36 GHz is similar to that of 
the $5_{-1}-4_0E$ lines, therefore we believe that the broad 36 GHz lines 
are also quasi-thermal. In analysing extended emission one need not 
distinguish between ``truly thermal" and ``quasi-thermal" lines, but negative 
excitation temperatures of some transitions may play a role in 
the appearance of compact maser spots in the corresponding cloud (\citet{sam98,kalen10}).

Statistical equilibrium calculations (e.g., ~\citet{cragg};) 
demonstrate that many Class I transitions, in
particular, $7_0-6_1A^+$, are inverted for a wide range of parameters
typical of Galactic molecular clouds. Therefore, the broad lines
detected at 44~GHz in S68N and Serpens CB2, as well as the broad line
in NGC~1333I2A, are most likely quasi-thermal lines.

\section{Conclusions}
A survey of young bipolar outflows in regions of low-to-intermediate-mass 
star formation has been carried out in two Class I methanol maser transitions, 
$7_0-6_1A^+$ at 44~GHz and $4_{-1}-3_0E$ at 36~GHz. As a result of the survey 
we detected narrow features at 44~GHz towards NGC~1333I2, NGC~1333I4A, 
HH25MMS, and L1157. One more maser candidate was detected at 36~GHz
towards the blue lobe of a bipolar outflow driven by a low-mass YSO in the NGC~2023
region. Flux densities of the lines detected at 44~GHz are no higher than 11 Jy
and their luminosities are about $10^{22}$~erg~s$^{-1}$, 
which is much lower than those of strong maser lines in regions of 
high-mass star formation. No emission was found towards 39 outflows. The upper limits 
for the non-detections are typically of the order of 3--5~Jy. Thus, new masers in 
regions of low-mass star formation should be searched for with a sensitivity 
of 1~Jy or better.

Observations at 44 GHz in 2004, 2006, and 2008 did not reveal a
significant variability of the masers in NGC~1333I4A, HH25MMS, and
L1157.

All masers at 44~GHz in these low-mass star formation regions were found in
clouds with methanol column densities of several times $10^{14}$~cm$^{-2}$ 
at linear scales of about 0.06 pc. Even higher methanol
column densities have been reported towards stronger
masers in regions of massive star formation. Therefore, the following
trend seems to exist: molecular clouds with methanol column densities less 
than $10^{14}$~cm$^{-2}$ cannot produce 44~GHz masers; clouds with methanol 
column densities $10^{14}-10^{15}$~cm$^{-2}$ can produce weak masers with 
luminosities about $10^{22}$~erg~s$^{-1}$; clouds with methanol column densities 
higher than $10^{16}$~cm$^{-2}$ can produce strong masers with luminosities 
$10^{24}-10^{25}$~erg~s$^{-1}$.

\section*{Acknowledgments}
The authors are grateful to the OSO staff for help during the observations. 
The work was partially
supported by the Russian Foundation for Basic Research (grants 
nos.~04-02-17057 and 07-02-00248) and the RAS Scientific Research 
Program "Extended Sources in the Universe". PH acknowledges partial 
support from NSF grant AST-0908901. The Onsala Space 
Observatory is the Swedish National
Facility for Radio Astronomy and is operated by Chalmers
University of Technology, G\"{o}teborg, Sweden, with financial
support from the Swedish Research Council and the Swedish Board
for Technical Development.

\clearpage\newpage
\onecolumn

\appendix
\section{List of non-detections at 44 and 36~GHz}

Table~\ref{negative} presents the list of non-detections at 44 and 36~GHz. 
The fifth column presents the LSR velocities that correspond to the center 
of the spectrometer bandwidth, the sixth column, the upper limits of flux 
densities at $3\sigma$ level. Note "hvb" means "high-velocity bullet", "p",
outflow with peculiar morphology; other notes are the same as in Table~\ref{gauss}.
\suppressfloats[t]
\begin{table*}
\centering
\begin{minipage}{160mm}
\caption{Non-detections at 44 and 36~GHz. }
\label{negative}
\begin{tabular}{|l|c|c|c|c|c|l|l|}
\hline
Source          &Line&R.A.      & DEC   &$V_{\rm LSR}$& $S_{\nu}$ &Notes &References\\
                &    &(J2000)   &(J2000)&(km~s$^{-1}$)&(Jy)       &      &\\
\hline
CB6             & 44 & 00:49:25.0 & +50:44:45.1 & -12.4 & 11.1& c & 22,28   \\ 
L1448IRS3       & 44 & 03:25:36.0 & +30:45:20.0 & -25.0 &  9.0& mo;c;b & 4,17,18,25   \\ 
L1448mm         & 44 & 03:25:38.8 & +30:44:05.0 &  67.0 &  9.3& mo;c;r & 4,17,18,23,25   \\ 
L1448mm         & 44 & 03:25:38.8 & +30:44:05.0 &   4.0 &  9.3& mo;c;r & 4,17,18,23,25   \\ 
L1448mm         & 36 & 03:25:38.8 & +30:44:05.0 &   4.0 &  2.4& mo;c;r & 4,17,23,25   \\ 
L1448           & 44 & 03:25:40.9 & +30:41:55.0 &  28.0 & 16.5& r & 4,17,25   \\ 
L1448           & 44 & 03:25:41.0 & +30:42:50.0 &  55.0 & 13.5& r & 4,17,25   \\ 
L1448           & 36 & 03:25:41.0 & +30:42:50.0 &  55.0 &  2.4& r & 4,17,25   \\ 
RNO15FIR        & 44 & 03:27:39.0 & +30:13:03.4 &   5.0 &  5.1& c & 13,14,18,23   \\ 
RNO15FIR        & 44 & 03:27:43.0 & +30:14:03.2 &   5.0 &  4.5& b & 13,14,18   \\ 
N1333I2A        & 44 & 03:28:48.0 & +31:14:55.0 &   2.9 &  3.9& b & 6,25,26   \\ 
N1333I2A        & 44 & 03:28:55.4 & +31:14:35.0 &   7.8 &  4.8& c & 6,18,23,25,26   \\ 
N1333I4A        & 44 & 03:29:06.5 & +31:12:18.5 &   7.0 &  3.9& b & 8,25,26   \\ 
IRAS03282       & 44 & 03:31:20.4 & +30:45:24.7 &   1.2 &  3.9& c & 3,5,18,25   \\ 
IRAS03282       & 44 & 03:31:30.3 & +30:43:34.2 &   1.2 &  4.2& b & 3,5,25   \\ 
IRAS03282       & 44 & 03:31:31.4 & +30:44:09.1 &   7.0 & 10.2& b & 3,5,25   \\ 
HH211           & 44 & 03:43:55.0 & +32:01:04.0 &  18.2 &  4.8& r & 24,35   \\ 
HH211           & 44 & 03:43:56.8 & +32:00:50.0 &   9.2 &  4.8& c & 24,35   \\ 
HH211           & 36 & 03:43:56.8 & +32:00:50.0 &   9.2 &  2.7& c & 18,24,35   \\ 
HH211           & 44 & 03:44:00.0 & +32:00:36.0 &   2.2 &  2.7& b & 24,35   \\ 
CB17            & 44 & 04:04:33.7 & +56:56:10.3 &  -4.7 &  4.2&   & 28     \\ 
L1489           & 36 & 04:04:43.0 & +26:18:56.9 &   7.0 &  3.0& c & 9,36   \\ 
IRAM04191       & 44 & 04:21:54.0 & +15:28:40.0 &   6.6 &  8.4& b & 29,30   \\ 
IRAM04191       & 44 & 04:21:57.0 & +15:29:46.0 &   6.6 &  3.0& c & 18,29,30   \\ 
L1527           & 44 & 04:39:53.9 & +26:03:10.4 &   6.0 &  5.4& c & 36   \\ 
L1527           & 36 & 04:39:53.9 & +26:03:10.4 &   6.0 &  2.7&   & 36   \\ 
CB26            & 44 & 04:59:52.4 & +52:04:45.1 &   5.8 &  3.3&   & 28   \\ 
IRAS05155       & 44 & 05:18:17.3 & +07:11:00.0 &  -1.6 & 10.5& c & 29   \\ 
OMC3            & 36 & 05:35:26.0 & -05:01:38.0 & 7.5 & 4.8& mo & 11,18,42   \\ 
OMC3            & 36 & 05:35:22.0 & -05:01:38.0 & 7.5 & 5.1& mo & 11,42   \\ 
OMC3            & 36 & 05:35:18.0 & -05:01:38.0 & 7.5 & 4.2& mo & 11,42   \\ 
OMC3            & 36 & 05:35:13.7 & -05:01:38.0 & 7.5 & 3.9& mo & 11,42   \\ 
OMC3            & 36 & 05:35:13.7 & -05:00:28.0 & 7.5 & 5.1& mo & 11,42   \\ 
OMC3            & 36 & 05:35:19.3 & -05:00:28.0 & 7.5 & 5.1& mo & 11,42   \\ 
OMC3            & 36 & 05:35:26.0 & -05:00:28.0 & 7.5 & 5.1& mo & 11,42   \\ 
OMC3            & 36 & 05:35:32.7 & -05:07:08.0 & 7.5 & 5.1& mo & 11,42   \\ 
OMC3            & 36 & 05:35:38.7 & -05:07:08.0 & 7.5 & 25.8& mo & 11,42   \\ 
OMC3            & 36 & 05:35:28.7 & -05:07:08.0 & 7.5 & 5.4& mo & 11,42   \\ 
OMC3            & 36 & 05:35:23.3 & -05:07:08.0 & 7.5 & 5.1& mo & 11,42   \\ 
OMC3            & 36 & 05:35:16.7 & -05:05:28.0 & 7.5 & 6.6& mo & 11,42   \\ 
OMC3            & 36 & 05:35:21.8 & -05:05:28.0 & 7.5 & 6.6& mo & 11,42   \\ 
OMC3            & 36 & 05:35:26.0 & -05:05:28.0 & 7.5 & 7.5& mo & 11,42   \\ 
OMC3            & 36 & 05:35:32.7 & -05:05:28.0 & 7.5 & 7.5& mo & 11,42   \\ 
IRAS05336 & 44 & 05:36:18.7 & -06:22:10.0 & 7.2 & 11.7& c & 18,23,38   \\ 
NGC2023 & 44 & 05:41:20.1 & -02:16:02.9 & 30.0 & 4.2& r & 37   \\ 
NGC2023 & 44 & 05:41:21.1 & -02:17:48.0 & 30.0 & 4.2& r & 37   \\ 
NGC2023 & 44 & 05:41:28.5 & -02:19:18.6 & -7.0 & 4.2& b & 37   \\ 
NGC2023 & 44 & 05:41:24.8 & -02:18:09.3 & 9.8 & 3.9& c & 18,37   \\ 
NGC2024FIR6 & 44 & 05:41:45.1 & -01:56:01.7 & 12.0 & 4.5& c & 10,18   \\ 
B35 & 36 & 05:44:29.8 & +09:08:53.7 & 11.7 & 2.1& r;b & 34,41   \\ 
HH212 & 44 & 05:43:49.0 & -01:04:10.0 & -10.0 & 7.5& r & 31   \\ 
HH212 & 44 & 05:43:51.4 & -01:02:53.0 & 1.7 & 5.1& c & 18,31   \\ 
HH212 & 44 & 05:43:54.0 & -01:01:30.0 & -10.0 & 6.9& b & 31   \\ 
\hline
\end{tabular}
\end{minipage}

\end{table*}

\begin{table*}
\centering
\begin{minipage}{160mm}
\contcaption{}
\begin{tabular}{|l|c|c|c|c|c|l|l|}
\hline
Source      &R.A.      & DEC   &Line&$V_{\rm LSR}$& $S_{\nu}$ &Notes &References\\
            &(J2000)   &(J2000)&    &(km~s$^{-1}$)&(Jy)       &      &\\
\hline
HH26M & 44 & 05:46:03.0 & -00:15:00.0 & 10.0 & 5.1& c & 14,21   \\ 
HH24MMS1 & 44 & 05:46:08.6 & -00:10:00.0 & 10.0 & 5.7&  & 14,21     \\ 
HH24MMS & 44 & 05:46:08.6 & -00:10:41.0 & 10.0 & 4.2& c & 14,18,21   \\ 
CB34 & 44 & 05:47:05.3 & +21:00:42.0 & 0.7 & 9.9& c & 22,28,43,44   \\ 
HH111B2 & 44 & 05:51:31.4 & +02:48:58.0 & -50.0 & 3.9& hvb & 32   \\ 
HH111B1 & 44 & 05:51:34.9 & +02:48:51.0 & -50.0 & 3.6& hvb & 32   \\ 
HH111O & 44 & 05:51:41.2 & +02:48:39.0 & 1.0 & 3.0& b & 32   \\ 
HH111MMS & 44 & 05:51:46.2 & +02:48:30.0 & 9.0 & 3.3& c & 18,32   \\ 
CB101 & 44 & 17:53:05.2 & -08:33:41.0 & 6.7 & 3.8&  & 22   \\ 
L483 & 44 & 18:17:33.2 & -04:39:44.1 & 9.0 & 3.9& r & 1,25,40   \\ 
L483 & 44 & 18:17:27.7 & -04:39:34.5 & 1.0 & 7.2& b & 1,18,25,40   \\ 
S68FIRS1 & 44 & 18:29:49.8 & +01:15:20.6 & 8.1 & 3.3& mo;c & 18,20,23   \\ 
SERP-SMM4 & 44 & 18:29:58.6 & +01:12:16.2 & 8.1 & 5.7& mo;b & 15,20   \\ 
SERP-SMM4 & 44 & 18:29:52.6 & +01:13:45.8 & 8.1 & 4.5& mo;b & 15,20   \\ 
SERP-SMM4 & 44 & 18:29:56.6 & +01:13:16.1 & 8.1 & 5.4& mo;c & 15,18,20   \\ 
L723K & 44 & 19:17:46.0 & +19:13:15.0 & 10.0 & 3.3& p;r & 25,29,2   \\ 
L723SE & 44 & 19:17:58.0 & +19:11:40.0 & 10.0 & 3.9& p;b & 29,2   \\ 
L723S1 & 44 & 19:17:50.0 & +19:11:30.0 & 10.0 & 3.3& p;r & 29,2   \\ 
CB199 & 44 & 19:37:10.2 & +07:36:50.0 & 8.4 & 3.9& c & 22   \\ 
CB205 & 44 & 19:45:21.3 & +27:50:40.0 & 8.0 & 3.0& c & 12,43,44   \\ 
L1157 & 44 & 20:39:04.0 & +68:04:45.0 & 15.0 & 1.8& r & 7   \\ 
L1157 & 36 & 20:39:04.2 & +68:03:30.0 & 15.0 & 1.5& r & 7   \\ 
L1157 & 44 & 20:39:04.2 & +68:03:30.0 & 15.0 & 3.0& r & 7   \\ 
CB230 & 44 & 21:17:39.4 & +68:17:31.9 & 2.7 & 2.4& c & 18,28,43,44   \\ 
CB232 & 44 & 21:37:11.3 & +43:20:36.0 & 12.6 & 3.3& c & 18,22,23,28,43,44   \\ 
NGC7129-FIRS1 & 44 & 21:43:20.0 & +66:08:00.0 & 0.0 & 5.1& r & 19   \\ 
NGC7129-FIRS2 & 44 & 21:43:01.7 & +66:03:25.0 & 0.0 & 2.4& mo;c & 18,19   \\ 
L1031 & 36 & 21:47:20.8 & +47:32:03.6 & 3.2 & 2.4& c & 16,34,41   \\ 
L1251A & 36 & 22:35:24.3 & +75:17:05.7 & -5.0 & 2.4& c & 23,33   \\ 
L1211-MMS1 & 44 & 22:47:02.2 & +62:01:31.0 & 14.0 & 3.3& c & 39   \\ 
L1211-MMS4 & 44 & 22:47:17.2 & +62:02:34.0 & -10.0 & 6.0& c & 18,39   \\ 
CepE & 44 & 23:03:13.0 & +61:42:59.0 & -11.2 & 7.2& r & 18,23,27   \\ 
CepE & 44 & 23:03:13.0 & +61:41:56.0 & -11.2 & 10.2& b & 23,27   \\ 
L1262A & 36 & 23:25:46.5 & +74:17:38.2 & 4.2 & 2.4& c & 18,41   \\ 
\hline
\end{tabular}
\end{minipage}
\flushleft

References:
1--Anglada et al. (1997); 2--Avery et al. (1990); 3--Bachiller et al. (1994); 4--Bachiller et al. (1995a);
5--Bachiller et al. (1995b); 6--Bachiller et al. (1998); 7--Bachiller et al. (2001); 
8--Blake et al. (1995); 9--Brinch et al. (2007); 10--Chernin (1996); 11--Chini et al. (1997);
12--Clemens et al. (1996); 13--Davis et al. (1997a); 14--Davis et al. (1997b); 15--Davis et al. (1999); 
16--Dobashi et al. (1992); 17--Dutrey et al. (1997); 18--Froebrich (2005); 19--Fuente et al. (2001); 
20--Garay et al. (2002); 21--Gibb \& Heaton (1993); 22--Gomez et al. (2006); 
23--de Gregorio-Monsalvo et al. (2006); 24--Hirano et al. (2006); 25--Kalenskii et al. (2007); 
26--Knee \& Sandell (2000); 27--Ladd \& Hodapp (1997); 28--Launhardt \& Henning (1997); 
29--Lee et al. (2002); 30--Lee et al. (2005); 31--Lee et al. (2006); 32--Lefloch et al. (2007); 
33--Meehan et al. (1998); 34--Myers et al. (1988); 35--O'Connell et al. (2005); 36--Ohashi et al. (1996);
37--Sandell et al. (1999); 38--Stanke \& Williams (2007,); 39--Tafalla et al. (1999); 
40--Tafalla et al. (2000); 41--Terebey et al. (1989); 42--Williams et al. (2003); 
43--Yun \&  Clemens (1992); 44--Yun \&  Clemens (1994);

\end{table*}

\label{lastpage}
\end{document}